\documentclass[useAMS,usenatbib]{mn2e}
\usepackage{graphicx,epsfig,amsmath,amssymb,amsbsy,natbib}

\usepackage{graphicx}
\usepackage{subfigure}
\title[A 100 kpc nebula associated with the ``Teacup'']{A 100 kpc nebula associated with the ``Teacup'' fading quasar}
\author[Villar Mart\'\i n et al.]{M. Villar-Mart\'{i}n$^{1,2}$, A. Cabrera-Lavers$^{3,4}$, A. Humphrey$^5$, M. Silva$^5$ \\
\newauthor   C. Ramos Almeida$^{4,6}$, J. Piqueras$^{1,2}$, B. Emonts$^7$ \\
$^1$Centro de Astrobiolog\'{i}a (CSIC-INTA), Carretera de Ajalvir, km 4, 28850 Torrej\'on de Ardoz, Madrid, Spain \\
$^2$Astro-UAM, UAM, Unidad Asociada CSIC, Facultad de Ciencias, Campus de Cantoblanco, E-28049, Madrid, Spain \\
$^3$GRANTECAN, Cuesta de San Jos\'e s/n, E-38712 , Bre\~na Baja, La Palma, Spain \\
$^4$Instituto de Astrof\'\i sica de Canarias, V\'\i a L\'actea s/n, E-38200 La Laguna, Tenerife, Spain \\
$^5$Instituto de Astrof\'{i}sica e Ci\^encias do Espa\c{c}o, Universidade do Porto, CAUP, Rua das Estrelas, PT4150-762 Porto, Portugal \\
$^6$Departamento de Astrof\'\i sica, Universidad de La Laguna (ULL), E-38205 La Laguna, Tenerife, Spain \\
$^7$National Radio Astronomy Observatory, 520 Edgemont Road, Charlottesville, VA 22903, USA}

\begin{document}

\date{Accepted ?.
      Received ?;
      in original form ?.}

\pagerange{\pageref{firstpage}--\pageref{lastpage}}
\pubyear{2013}

\maketitle

\label{firstpage}

\begin{abstract}

We report the discovery of a $\sim$100 kpc ionized nebula associated with the radio quiet type 2 quasar (QSO2) nicknamed the ``Teacup'' ($z$=0.085).  The giant nebula is among the largest known around active galaxies at any $z$. We propose that it is part of the circumgalactic medium (CGM)  of the QSO2 host, which has been populated with tidal debris  by galactic interactions. This rich gaseous medium has been rendered visible due to the illumination by the powerful active nucleus (AGN).    Subsolar abundances ($\sim$0.5$Z_{\rm \odot}$) are tentatively favored by AGN photoionization models. We also report the detection of coronal emission (Fe$^{+6}$) from  the NE bubble, at $\sim$9 kpc from the AGN. The detection of coronal lines at such large distances from the AGN and the [NII]$\lambda$6583/H$\alpha$, [SII]$\lambda\lambda$6716,6731/H$\alpha$, [OI]$\lambda$6300/H$\alpha$ optical emission line ratios of the giant nebula   are  consistent with   the fading quasar scenario proposed by \cite{gag14}.  The fading rate appears to have been faster in the last $\sim$46,000 yr.  Deep wide field integral field spectroscopy of  giant nebulae around powerful AGN such as the ``Teacup's''  with instruments such as MUSE on VLT  opens up a way to detect and study the elusive material from the CGM around massive active galaxies thanks to the illumination by the luminous AGN. 
 
\end{abstract}


\section{Introduction}

The ``Teacup''  (SDSS J143029.88+133912.0 at $z=$0.085) is a radio quiet type 2 quasar (QSO2), whose nickname comes from the peculiar morphology of the extended ionized gas. It shows a loop-shaped  emission line structure reminiscent of a ``handle''   extending up to $\sim$12 kpc  NE of the active galactic nucleus (AGN) (Keel et al. \citeyear{keel12}, Gagne et al. \citeyear{gag14}, Ramos Almeida et al. \citeyear{ram17}). This object has been subject of an intensive study  by different groups for two main reasons. On one hand, the modeling of the emission line spectra with AGN photoionization models has led to the conclusion that it is  a fading quasar
(Gagne et al.  \citeyear{gag14}, Keel et al. \citeyear{keel17}). The luminosity of the active nucleus appears to have dropped by $\sim$2 orders of magnitude in the last $\sim$46,000 years.  On the other hand, the system has been proposed  to be the scenario of a giant outflow generated by an AGN wind or induced by a $\sim$1 kpc   radio jet whose effects are noticed up to $\sim$10-12 kpc from the AGN and may be responsible for the  bubble-like morphology inspiring its nickname (Harrison et al. \citeyear{har15}). 
A detailed description of the general properties of  the ``Teacup'' can be found in this paper. 

We present here  new results based on optical long slit spectroscopic data obtained with the Spanish Gran Telescopio Canarias (GTC). We analyze and interpret the properties (size, kinematics, line ratios) of a newly discovered giant reservoir of ionized gas associated with the ``Teacup'', which extends for more than 100 kpc. New results related to the coronal emission from the NE ionized bubble are also presented. The paper is organized as follows: we describe  briefly the observations and data reduction in Sect. \ref{obs}. Results are presented in Sect. \ref{results} and discussed   in Sect. \ref{discuss} and  summarized  Sect. \ref{Conc}.

We adopt $H_{0}$=71 km s$^{-1}$ Mpc$^{-1}$, $\Omega_{\Lambda}$=0.73 and
$\Omega_{m}$=0.27.  This gives an
arcsec to kpc conversion 1.58 kpc arcsec$^{-1}$ at $z$=0.085. 
\begin{figure*}
\includegraphics{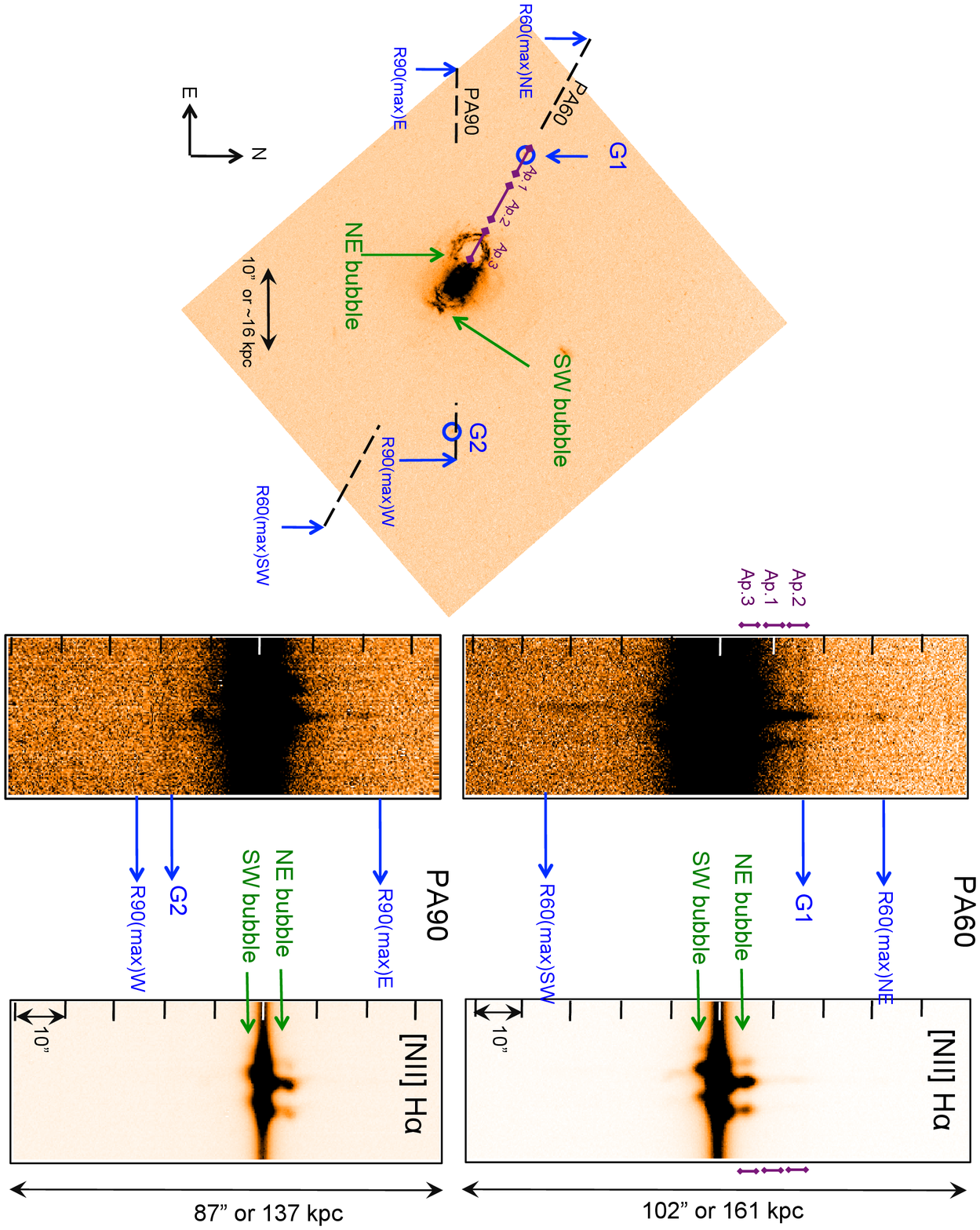}
\vspace{5.1in}
\caption{HST  WFC3  FR716N  image of the ``Teacup''  containing the H$\alpha$+[NII] lines. We highlight prominent emission line features such as the NE and SW bubbles. $R60(max)_{\rm NE}\sim$53 kpc, $R60(max)_{\rm SW}\sim$58 kpc, $R90(max)_{\rm E}\sim$40 kpc and $R90(max)_{\rm W}\sim$31 kpc  are the maximum radial distances of the H$\alpha$ emission measured with the spectra along the PA60 and PA90 slits in directions NE,  SW, E and W of the QSO2 nucleus respectively. The nebula is giant in comparison with the  bubbles. The approximate location and size of apertures 1, 2 and 3 (Ap. 1, Ap. 2, Ap. 3) used to study the emission line spectra of the giant nebula and the NE bubble are also shown (see text). See the electronic edition for colour version of the figures.}
\label{fig:hst1}
\includegraphics{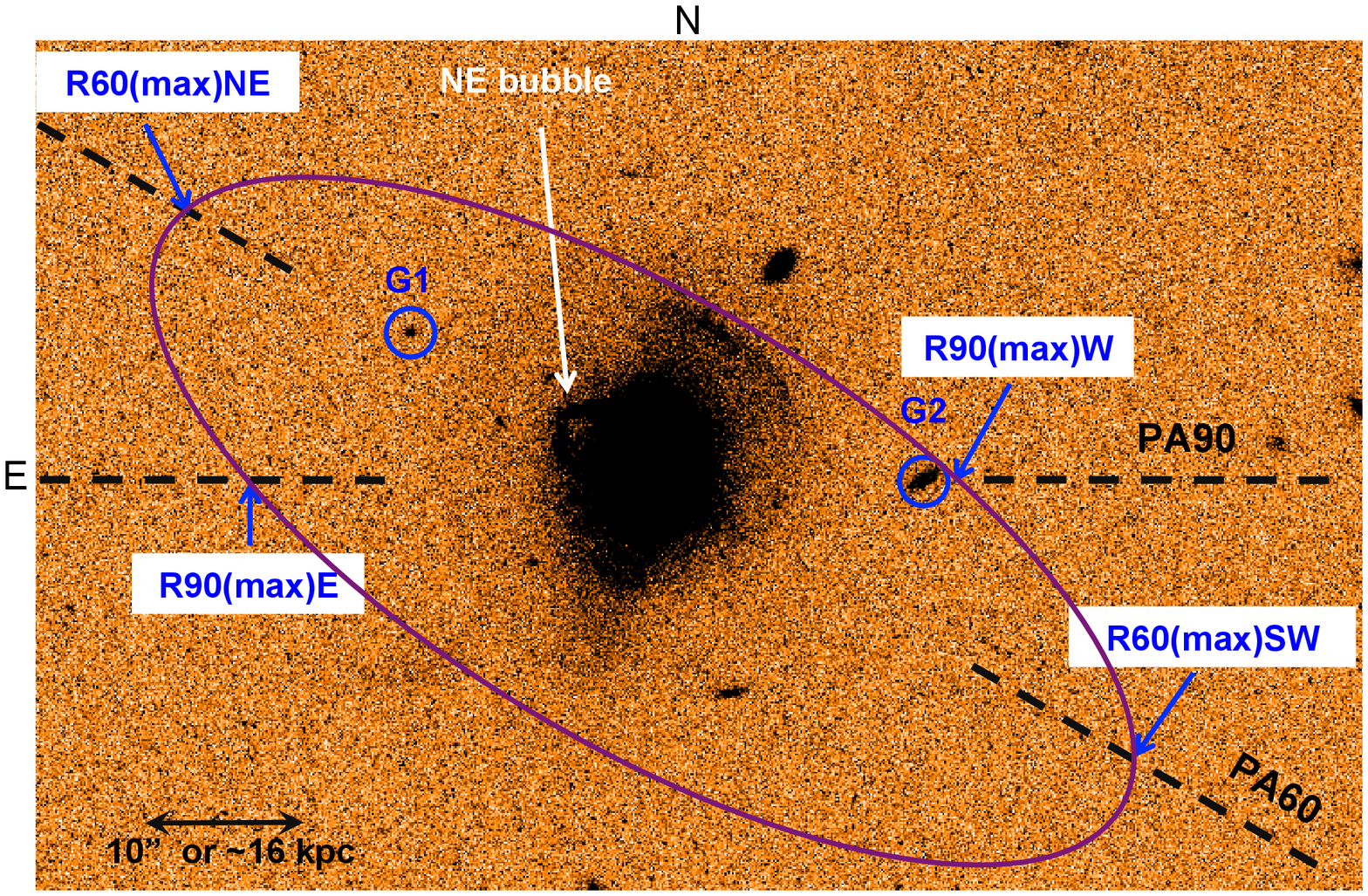}
\vspace{2.8in}
\caption{HST WFC3 F763M continuum image. We show again the striking dimensions of the giant gaseous reservoir around the Teacup. The purple elliptical area shows a possible  geometry of the  nebula (for instance, if settled in a rotating disk), adjusted in angle and size to fit the radial sizes of H$\alpha$ in different directions from the QSO2 along PA60 and PA90.}
\label{fig:hst2}
\end{figure*}

\section{Observations and data reduction}
\label{obs}
Long slit spectroscopic observations were performed on February 26th 2017 (programme GTC13-16B) in visitor mode with  the optical imager and long slit spectrograph OSIRIS\footnote{http://www.gtc.iac.es/en/pages/instrumentation/osiris.php} mounted on the 10.4m GTC. 
The    R2500R volume-phased holographic grating (VPH) was used, which provides a spectral coverage of  5575-7685 \AA\ with   dispersion  1.04 \AA\ pixel$^{-1}$. A 1.5 arcsec wide slit was used at two different orientations (Fig. \ref{fig:hst1}): position angle PA 60$^\circ$ (N to E), along the main axis of the NE bubble and  PA 0$^\circ$, in the E-W direction along the South edge of this bubble. The spectral resolution FWHM$_ {\rm inst}$ measured from several prominent sky lines is  5.28$\pm$0.14  \AA\ (223$\pm$6 km s$^{-1}$ at the the observed H$\alpha$ wavelength). The pixel scale is 0.254 arcsec pixel$^{-1}$. 

Eight 300 sec spectra were obtained at each orientation to complete a total exposure time of 2400 sec per PA. Shifts   were applied in the slit direction of 20" between consecutive exposures for both a better fringing correction and a more accurate background subtraction. The seeing FWHM during the observations was 1.4$\pm$0.1 arcsec (FWHM), as measured from several stars in the acquisition images. The spectra were reduced and flux calibrated following standard procedures (see Villar Mart\'\i n et al. \citeyear{vm17} for details). 

\subsection{Additional data}

We have also used two archive HST images of the ``Teacup'' (program 12525, principal investigator W.C. Keel; Keel et al. \citeyear{keel15}).  The emission line+continuum image was obtained on June 14th 2012 with the Advanced Camera for Surveys (ACS)  and the WFC detector. This provides a pixel scale of 0.05 arcsec pixel$^{-1}$. 
The  FR716N ramp filter was used, which covers the $\sim$6850-7470 \AA\  spectral range ($\sim$6313-6885 \AA\ rest frame for the ``Teacup''). It thus includes H$\alpha$+[NII]$\lambda\lambda$6548,6583 and [SII]$\lambda\lambda$6716,6731  and emission line structures will appear very prominently. 

The continuum image was obtained on June 13th 2012 with the Wide Field Camera 3 (WFC3).
The   F763M medium band filter was used in the UVIS channel, which provides a pixel scale of 0.04 arcsec pixel$^{-1}$. The filter  covers the $\sim$7243-8036 \AA\ range  ($\sim$6675-7406 \AA\  rest frame). The  image is  dominated by continuum emission. Some non-negligible contribution of  lines such as [SII]$\lambda\lambda$6716,6731 may be expected in emission line dominated structures.

\section{Results}
\label{results}

\subsection{The giant ionized gas reservoir}
\label{nebula}

We show in Fig. \ref{fig:hst1} the HST  FR716N  image of the ``Teacup''.  Emission line structures such as the ionized bubbles  appear very prominently.  The two GTC-Osiris slit positions at PA60 and PA90 are over-plotted. The  2D spectra of  [NII]$\lambda\lambda$6548,6583+H$\alpha$  along both PA are also shown with different contrasts to highlight the low and high surface brightness structures. Several  features are also identified. G1 (PA60) is an emission line galaxy at $z$=0.317. G2 (PA90) is also a  galaxy. The detection of a line at  5851 \AA\ (probably [OII]$\lambda$3727) and the tentative detection of another one at  $\sim$7635 \AA\ (probably H$\beta$) suggest  G2 is at $z=$0.570.

The 2D spectra of the ``Teacup'' reveal line emission across giant extensions along both PA. 
H$\alpha$  is detected up to $R60(max)_{\rm NE}\sim$53 kpc and 
$R60(max)_{\rm SW}\sim$58 kpc to the NE and SW respectively from the continuum centroid along PA60.  It
 extends up to $R90(max)_{\rm E}\sim$40 kpc to the E along PA90. Detection is confirmed up to  $R90(max)_{\rm W}\sim$31 kpc to the W. Fainter H$\alpha$ is tentatively detected up to $\sim$63  kpc, although this needs confirmation with deeper data. 
 
 We have thus discovered a giant reservoir of gas ($>$100 kpc total extension) associated with the ``Teacup'', which extends far beyond the  radio and optical bubbles, each  one  having radial sizes of $\sim$10-12 kpc.  The long slit and integral field spectroscopic studies of \cite{gag14} and \cite{har15}  already revealed the existence of gas beyond the bubbles.  However, due to the shallowness of the data and/or the small field of view of the instruments, emission was confirmed only up to $\sim$11 arcsec or 17.4 kpc from the AGN\footnote{The comparison of Gagne et al. (2014) velocity curve along PA95 in their Fig. 8  with our own data leads us to conclude
  that their  spatial axis is  in pixels instead of arcsec.}. 
  
We show in Fig. \ref{fig:hst2}  the FR763M HST image. Low surface brightness asymmetric  continuum features (tidal tails, shells) are appreciated across several 10s of kpc. 
They are the remnants of the merger of   a small, dynamically cold system with the bulge dominated QSO2 host (Keel et al. \citeyear{keel15}).
The maximum radial extensions of the H$\alpha$ emission along the long slit spectra are  indicated with blue arrows.  
  Despite the incomplete spatial coverage in two spatial dimensions, the huge extension along both PA  
and the rather  symmetric radial sizes at both sides of the AGN suggest that the ``Teacup'' is embedded in a giant gaseous reservoir. 
The purple ellipse  shows a possible  geometry of the  nebula, adjusted in angle and size to fit the radial sizes of the H$\alpha$ extension in different directions from the QSO2 along PA60 and PA90.  The ionized bubbles  are comparatively tiny.

\subsubsection{Ionization mechanism}
\label{ionization}

 We have extracted two 1D spectra from the highest surface brightness regions of the nebula along PA60, beyond the NE bubble (Fig. \ref{fig:hst1}).  Aperture 1 (Ap. 1) is centered at 11.5 arcsec (18.2 kpc) from the QSO2 continuum centroid and is 4.8 arcsec wide (7.6 kpc). The G1 galaxy is within this aperture, but no emission line contamination is expected by it
 because the redshift $z$=0.317 is very different. Aperture 2 (Ap. 2) is centered   at 16.6 arcsec (26.2 kpc)  and is 3.6 arcsec wide (5.7 kpc). [NII]$\lambda\lambda$6548,6583, H$\alpha$ and [SII]$\lambda\lambda$6716,673 are detected. [OI]$\lambda$6300 is affected by a sky residual.

\begin{figure*}
\includegraphics{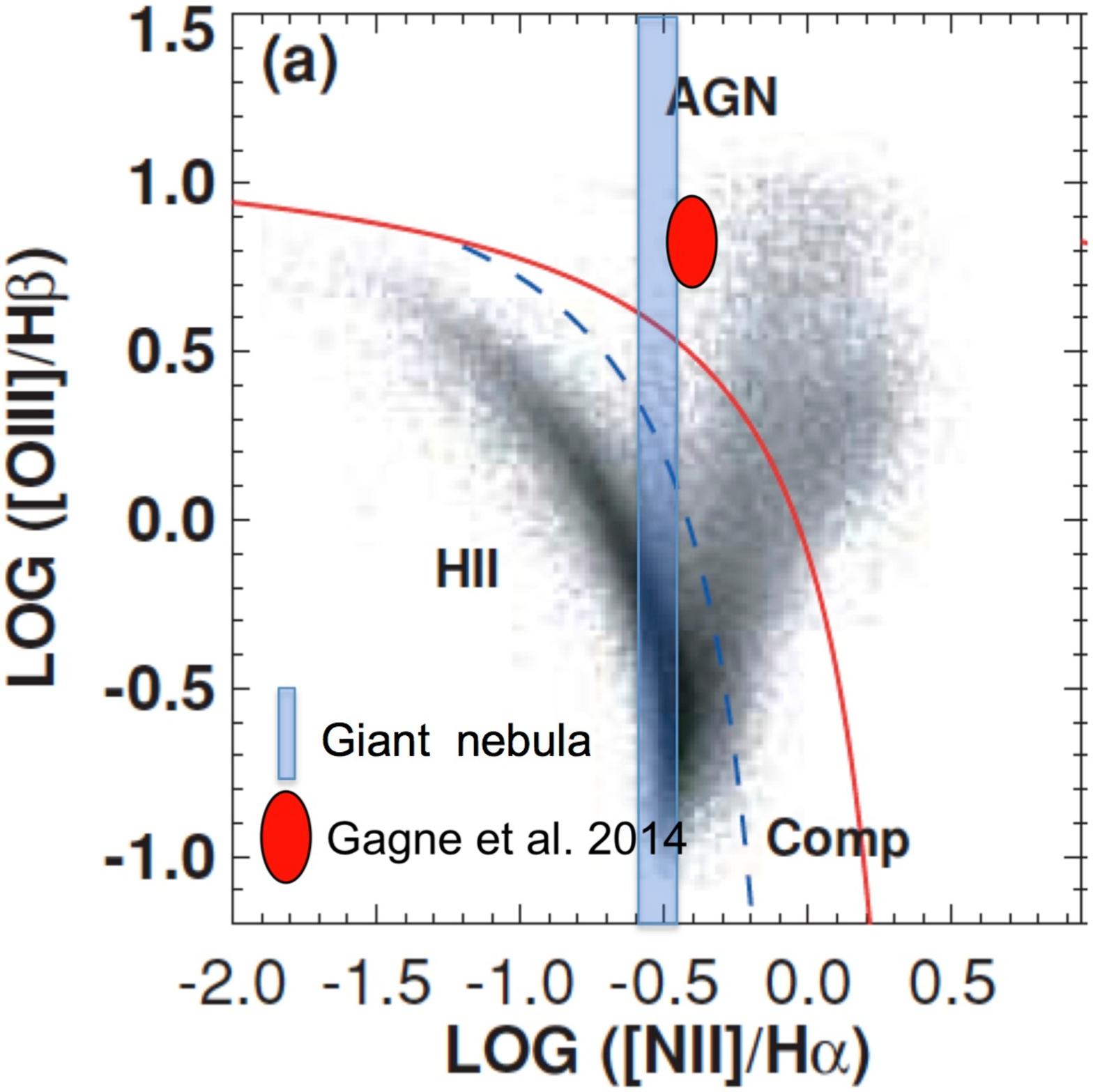}
\includegraphics{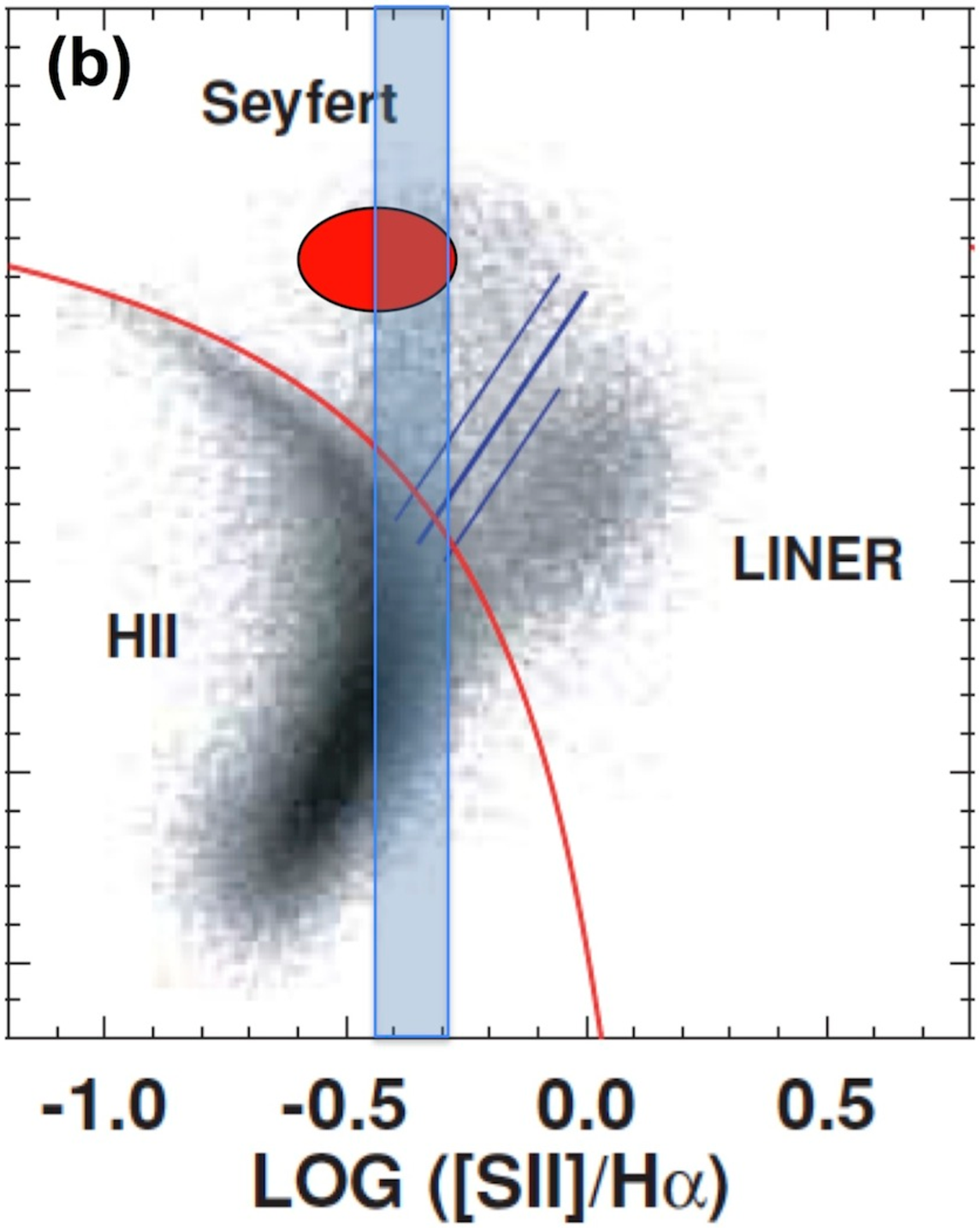}
\includegraphics{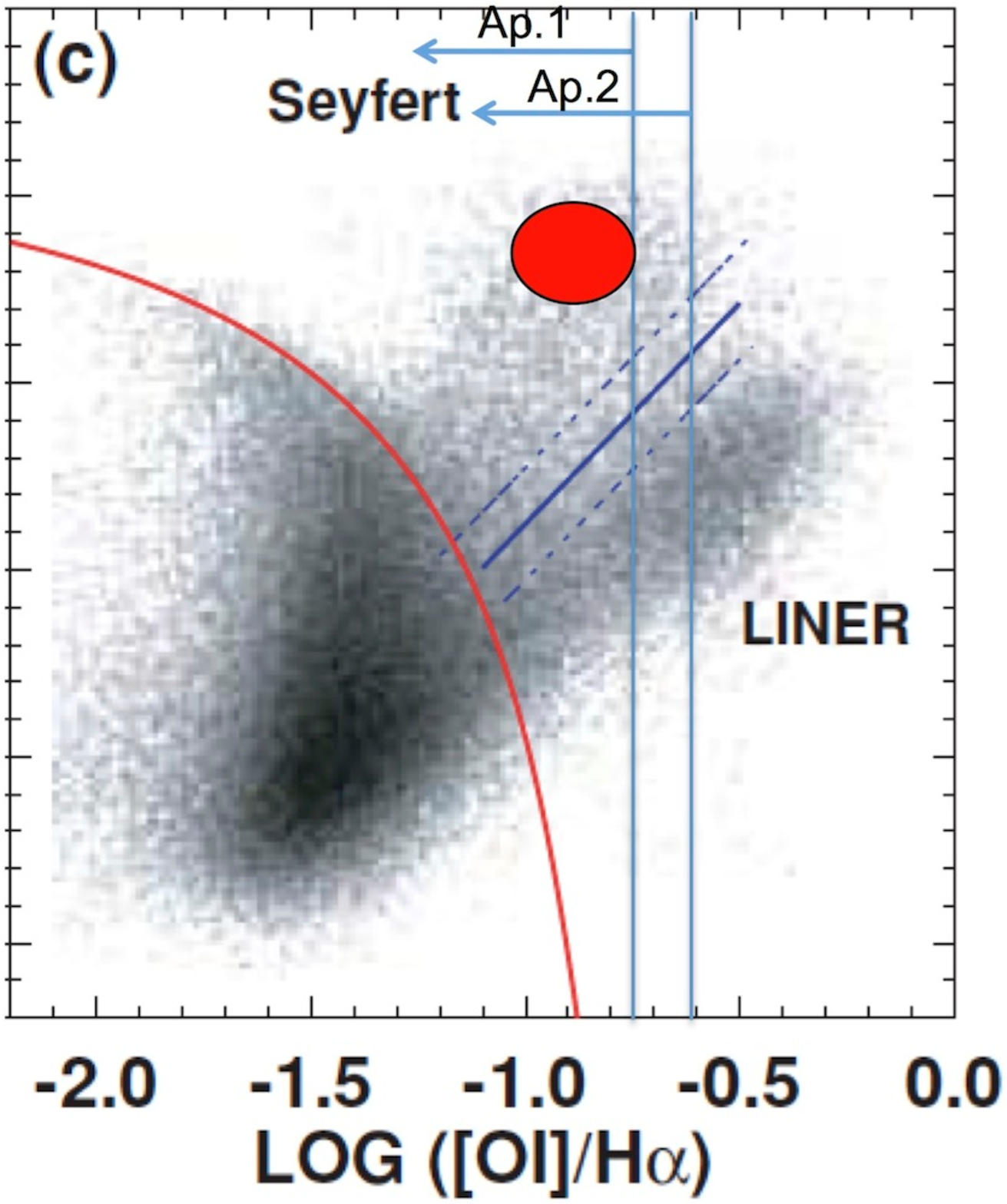}
\vspace{3.0in}
\caption{BPT (Baldwin et al. 1981) diagnostic diagrams taken from Kewley  et al. (2006).  Since the [OIII]/H$\beta$ ratio
is not available for the ``Teacup'' nebula , only the location on the horizontal axis is constrained. The blue shadowed areas represent the range of values spanned by Ap. 1 and Ap. 2 (see text) including the uncertainties. It is not possible to discriminate the ionization mechanism from these line ratios alone. The red filled ellipses show the approximate area covered by the line ratios measured by Gagne et al. (2014) at different locations closer to the AGN, such as the NE and SW bubbles and the nucleus. The fact that the nebula presents line ratios that are consistent with those spanned by the red ellipses suggests that the same mechanisms (AGN photoionization) excites the gas.}
\label{fig:diag}
\end{figure*}

The  [NII]$\lambda$6583/H$\alpha$, [SII]$\lambda\lambda$6716,6731/H$\alpha$ ratios and
[OI]$\lambda$6300/H$\alpha$ upper limits   are shown in Table \ref{tab:tabgiant} for both apertures. It is not possible to discriminate the ionization mechanism from these values alone (see Fig. \ref{fig:diag}). On the other hand, they  are consistent with   those measured by \cite{gag14} for gas at numerous locations closer to the AGN, including different positions across the NE and SE bubbles  (see Fig. \ref{fig:diag}, red ellipses). This suggests that the same ionizing mechanism excites the gas of the giant nebula. This mechanism is AGN photoionization (Gagne et al. \citeyear{gag14}; see also Keel et al. \citeyear{keel17}) 

The  [SII] doublet falls between two prominent sky emission lines so that sky residuals do not pose a problem to measure the doublet flux ratio.  Using  the Ap. 1 spectrum, for which the lines are stronger, we obtain [SII]$\frac{\lambda 6716}{\lambda 6731}$=1.33$\pm$0.10.  Assuming an electron temperature in the range $T_ {\rm e}\sim$ 10,000-20,000 K, then $n\la$ 240 cm$^{-3}$ (Osterbrock \& Ferland \citeyear{ost06}). 

\begin{table*}
\centering
\begin{tabular}{lllllll}
\hline
Aperture & Distance to AGN & Width  & $\frac{[NII]\lambda 6583}{H\alpha}$  &   $\frac{[OI]\lambda 6300}{H\alpha}$  &  $\frac{[SII]\lambda\lambda 6716,6731}{H\alpha}$  \\ 
	& 	 kpc 	(NE) & kpc   & & &    \\ \hline
Ap. 1 	&  18.2	&  7.6  & 0.31$\pm$0.03 & $\la$0.18 & 0.47$\pm$0.04 \\
Ap. 2 	&  26.2	&  5.7  & 0.27$\pm$0.03 & $\la$0.25 & 0.40$\pm$0.04  \\
\hline
\end{tabular}
\caption{Line ratios for apertures 1 and 2 (see text) along PA60 on the giant nebula  at $\sim$18 and $\sim$26 kpc  from the AGN well beyond the NE bubble. The width quoted in the 3rd column is the physical size of the apertures in kpc. }   
\label{tab:tabgiant}
\end{table*}

\subsubsection{Kinematics of the giant nebula }
\label{kinem}

\begin{figure*}
\includegraphics{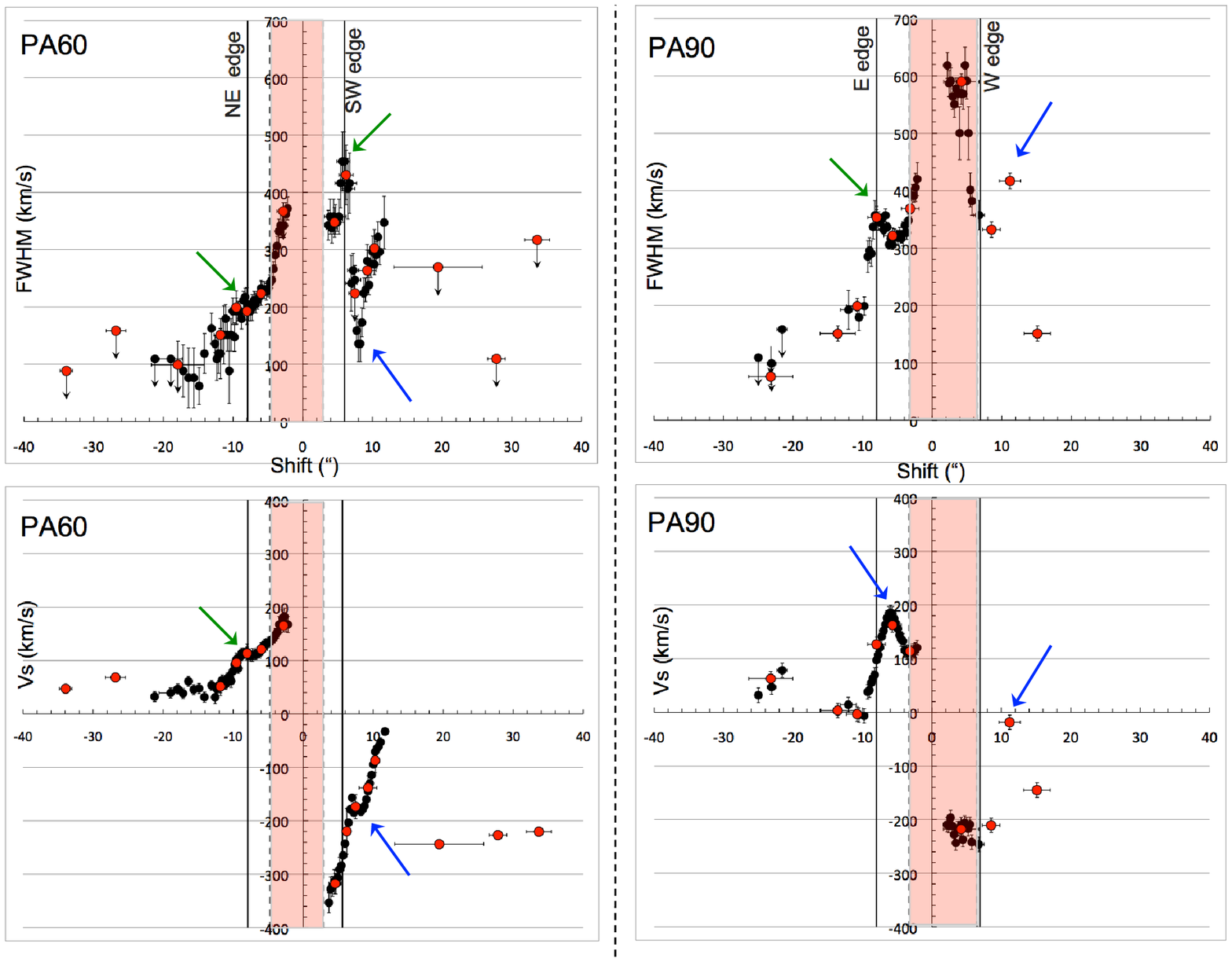}
\vspace{5.5in}
\caption{Kinematics of the extended gas along PA60 (left) and PA90 (right). The red shadowed areas indicate regions in which contamination by the nuclear emission is significant. The vertical black solid lines mark the approximate location of the edges of the radio bubbles (Harrison et al. 2015). Red solid circles correspond to measurements using larger apertures to increase the signal to noise ratio. The aperture sizes are indicated by the size of the horizontal bars. The green arrows mark  kinematic changes  identified at or very near  the bubble edges. The blue arrows mark prominent kinematic changes not clearly associated with the  bubbles.}
\label{fig:kinem}
\end{figure*}

We show in Fig. \ref{fig:kinem} the spatial variation of the FWHM (corrected for instrumental broadening) and the velocity shift $V_{\rm s}$ of H$\alpha$ relative to the  nuclear emission line. The red shadowed areas mark the regions where contamination by the seeing smeared nuclear emission is expected to be significant relative to the extended emission (Villar Mart\'\i n et al. \citeyear{vm16}). Due to the complexity of the heavily blended nuclear [NII]+H$\alpha$  and the fact that we are mostly interested on the giant nebula and its properties in comparison with the bubbles, we do not show measurements within the inner areas where the emission is clearly dominated by the nucleus.
The edges of the radio bubbles (Harrison et al. \citeyear{har15}) are marked with vertical black solid lines. 
 
 The H$\alpha$ kinematics  become more quiescent  at both sides of the AGN in the outer parts of the giant nebula. At $r>$12 arcsec (19 kpc) the lines are narrower (in general unresolved with FWHM$\la$150 km s$^{-1}$) and the $V_{\rm s}$ curve becomes almost flat. This is very clear along PA60 and can be hinted  along PA90. 
At smaller radii, the lines are broader (FWHM  $\sim$several$\times$100 km s$^{-1}$) and present a more complex velocity field with steep  $V_{\rm s}$ profiles. 

The kinematics of the gas within the bubbles have been studied by different authors. \cite{gag14} proposed a rotational pattern dominating the motions within the inner $\sim$5 arcsec with some disturbance  possibly due to a merger. Small kinematic changes at the NE bubble edge (named ``the handle'' by the authors) also revealed the localized action of another mechanism that they propose to be an outflow. Other authors (Harrison et al. \citeyear{har14}, \citeyear{har15},  Keel et al. \citeyear{keel17}, Ramos Almeida et al. \citeyear{ram17})  have   interpreted the motions in terms of the action of the giant outflow inflated by the AGN wind or the radio structures, acting across the whole volume covered by  the expanding bubbles.  \cite{keel17} claim that asymmetric [O III] profiles are present across the NE bubble, reaching maximum velocities of $\pm$1000 km s$^{-1}$.   \cite{ram17}  also report tentative detection of a very broad Pa$\alpha$ component of FWHM$\sim$3000 km s$^{-1}$ at different locations across the bubbles. Such extreme kinematics would strongly support the outflow scenario.  

We do not detect such extreme motions (neither do Harrison et al. \citeyear{har15}). We find turbulent gas (up to FWHM$\sim$450 km s$^{-1}$) across the bubbles (as found by Ramos Almeida et al. \citeyear{ram17} as well),  but  the kinematics are not extreme.  We consider such  FWHM  values indicative of kinematic turbulence because even in the most dynamically   disturbed mergers with signs of AGN activity, the extended non outflowing ionized gas shows typical FWHM$<$ 250 km s$^{-1}$  (Bellochi et al. \citeyear{bel13}). 

We also find that at the edges of both bubbles the gas presents the following  kinematic changes compared with the gas within and beyond those locations (green arrows in Fig. \ref{fig:kinem}): a) a clear broadening of the lines at  the edge of  SW bubble along PA60 (Fig. \ref{fig:kinem}, top left panel),  b) a clear broadening of the lines at the edge of the NE bubble to the E along PA90, c) a slight broadening of the lines at the edge of the NE bubble to the E (top right panel) and d) a slight change of  $V_{\rm s}$ at the edge of the NE bubble along PA60 (bottom left panel). 
These results  support that the expanding bubbles have an impact (at least local) on the underlying kinematic pattern. 

However, a potential problem for the outflow scenario is that turbulent gas and prominent kinematic changes are also identified well beyond the bubble edges (blue arrows in Fig. \ref{fig:kinem}). This is  specially clear for gas beyond the SE bubble. At $\sim$8 arcsec from the AGN along PA60 the line FWHM  starts to increase sharply from $\sim$140 km s$^{-1}$ up to $\sim$350 km s$^{-1}$ at $\sim$12 arcsec (Fig. \ref{fig:kinem}, top left panel). This  FWHW occurs far beyond the SW bubble edge, at $\sim$6 arcsec or 9.5 kpc.  
 Along PA90, at $\sim$11 arcsec  to the W  the lines are significantly broader (FWHM$\sim$420 km s$^{-1}$) and much less blueshifted ($V_{\rm s}\sim$-20 km s$^{-1}$) than neighboring regions ($V_{\rm s}\la$-145 km s$^{-1}$). This gas is $\sim$4.7 arcsec or 7.4 kpc beyond the SW bubble edge. Such broad lines are rather extreme at such large distance from the galaxy nucleus ($\sim$19 kpc). 
  
  Therefore, turbulent gas exists far beyond the extension of the outflow traced by the bubbles. 
 It is thus possible,  that the gas motions within the inner $\sim$12 arcsec are dominated by a mechanism other than the outflow, while this produces some localized perturbance  at the edges of the expanding bubbles.  It is not clear what that mechanism may be. 
 The redistribution of gas by mergers/interactions may play a  role. This is supported by the fact that the highest surface brightness stellar shells and tidal features seen in the continuum HST image (Fig. \ref{fig:hst2})  are distributed within $r\la$12 arcsec from the AGN.   However,  as mentioned above,  it is not clear that this mechanism alone can produce the large FWHM measured. 
 
  Alternatively, it is possible that the outflow reaches larger distances than mapped by the radio bubbles and/or that the effects from previous outflow episodes are still visible. This would not be surprising, since the radio emission shows  mostly the working surface of the actual radio source, where the radio plasma currently interacts more strongly with the interestellar and/or intergalactic medium. Deeper or low frequency data may  trace  emission further out. Indeed, this is supported by the [OIII]$\lambda$5007 morphology  of  \cite{har15} (see their Fig. 5), which shows faint ionized gas beyond  the radio bubbles edges which and roughly circumscribing them. 

 Therefore, there is not a definite explanation for the motions  of the gas at $\la$12 arcsec  from the AGN. A combination of the two mechanisms mentioned above may be at work.  Disentangling this issue would be of great interest to understand the real impact of the radio  or AGN induced outflow across its environment.  
 
 A relevant result is that the motions of the outer nebula  ($\ga$12 arcsec) are detached from the kinematic pattern of the inner regions.  

\subsection{Coronal emission from the NE bubble}
\label{coronal}

The   kinematic, ionization, and morphological properties of the ionized bubbles (specially the NE one) have been studied in great detail in the literature (Keel et al. \citeyear{keel12},\citeyear{keel15},\citeyear{keel17}, Gagne et al. \citeyear{gag14}, Harrison et al. \citeyear{har14},\citeyear{har15}, Ramos Almeida et al. \citeyear{ram17}). The new result we report here is the detection of coronal emission from the NE bubble.

\begin{figure*}
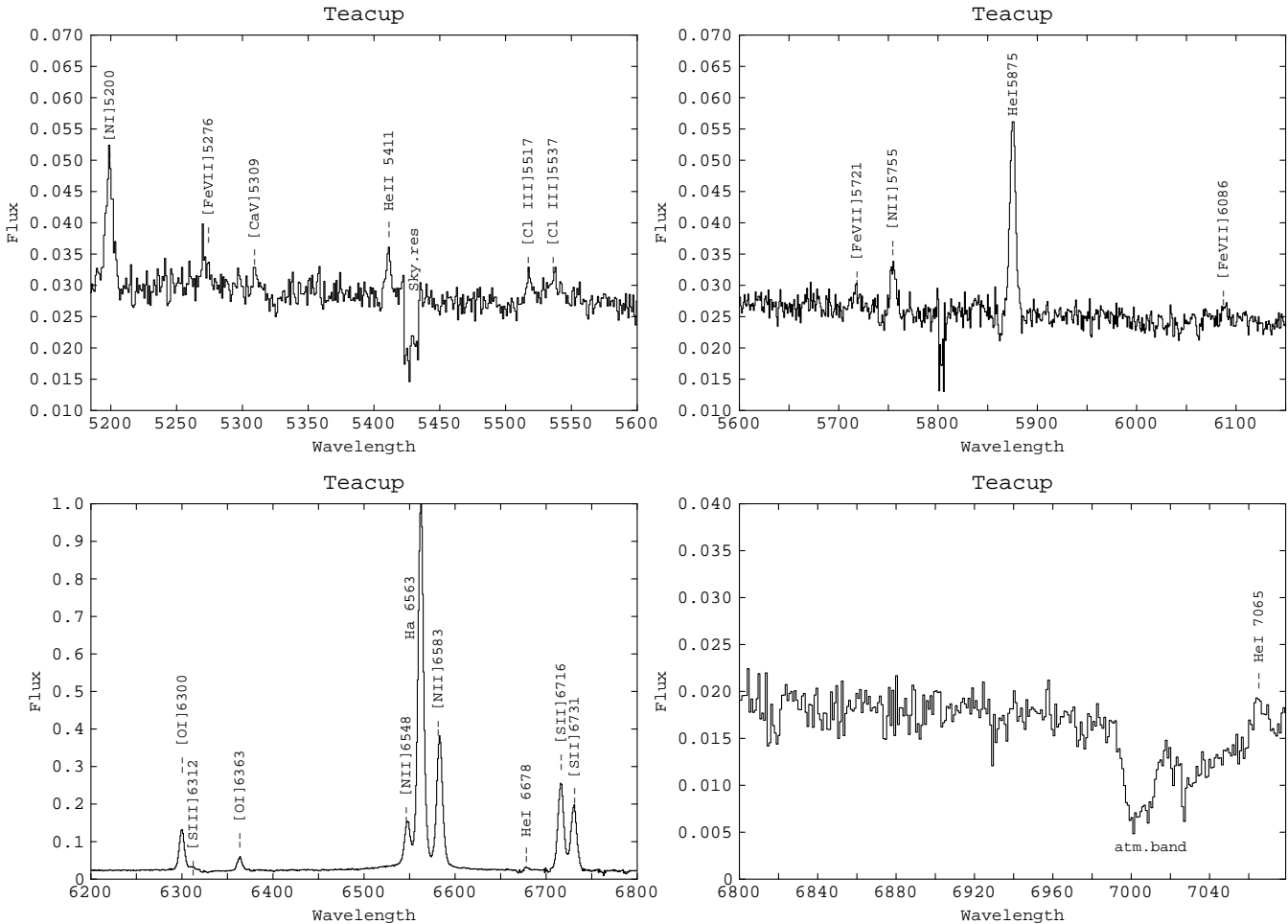

\includegraphics{portion1.ps}
\includegraphics{portion2.ps}
\vspace{2.7in}
\includegraphics{portion3.ps}
\includegraphics{portion4.ps}
\vspace{2.5in}
\caption{Spectrum of the NE bubble split in four  spectral windows with different flux scales to highlight some of the weakest lines.
Wavelength (rest-frame) in \AA. The flux in all spectra has been normalized to the peak of H$\alpha$. Several coronal lines are detected: [FeVII]$\lambda$5276, $\lambda$5721, $\lambda$6086.}
\label{fig:specbubble}
\end{figure*}

 We extracted  a 1D spectrum of this region from an aperture selected to reach a compromise between S/N for the measurement of faint emission lines and minimizing the contamination from the QSO2 nuclear emission due to seeing smearing (Villar Mart\'\i n et al. \citeyear{vm16}).   This aperture is centered at 5.7 arcsec (9.0 kpc)  NE from the QSO2 continuum centroid along PA60 and is 3.8 arcsec (6.0 kpc) wide (Ap. 3 in Fig. \ref{fig:hst1}). \cite{gag14} already showed that both the NE  and SW bubbles  emit  rich emission line spectra. The larger aperture of the GTC has allowed to identify fainter lines undetected before (see Table \ref{tab:bubblelines} and Fig. \ref{fig:specbubble}), most interestingly,  several coronal lines  (CL) emitted by Fe$^{+6}$ ([FeVII]$\lambda$5276,$\lambda$5721 and $\lambda$6086). Based on the [SiVI]$\lambda$1.963 $\mu$m line, \cite{ram17}  confirmed the detection of an extended coronal outflow of size $\sim$1 kpc almost coincident with the inner radio jet. The GTC spectrum shows that also the NE bubble emits CL, which therefore has a very broad range of ionization species, neutral (O$^0$),  ionized (He$^+$, H$^+$, O$^{+3}$, etc) and coronal (Fe$^{+6}$). 
 [FeVII]$\lambda$5721   is detected also  in the PA90 bubble spectrum. This shows that coronal emission occurs at different locations across the bubble.

The coronal lines are possibly blueshifted relative to lower ionization lines (Table \ref{tab:bubblelines}), although they are  faint and higher S/N spectra would be necessary to determine the precise redshift more accurately.  If confirmed, this would reveal kinematic substructure in the bubbles dependent on the gas ionization level.

\begin{table*}
\centering
\begin{tabular}{lllllll}
\hline
(1) & (2) & (3) & (4) & (5)  \\
Species & $\lambda_{\rm air}$&  $\lambda_{\rm obs}$ & $\Delta(V)$ & 100$\times$$\frac{Flux}{Flux(H\alpha)}$  \\   
& (\AA) & (\AA)  & km s$^{-1}$  &  &  \\ \hline
~[NI]  	& 5197.9+5200.4	&  5643.9$\pm$0.5 & -24$\pm$24   & 1.88$\pm$0.02   \\
~[Fe VII]$^{*}$ & 		5276.4	& 5724.1$\pm$0.7 & -197$\pm$28 &   0.50$\pm$0.05  \\
 ~[CaV] 	&    	5309.1	& 5764.5$\pm$0.6	&   60$\pm$28  &  0.32$\pm$0.06		\\
 ~HeII 	&    	5411.5	& 5874.5$\pm$0.2	  &   0$\pm$15 &  0.67$\pm$0.07	 	\\
~[Cl III]	&   5517.7		& 	5990.7$\pm$0.4 &  44$\pm$21   &  		0.36$\pm$0.08  \\  
~[Cl III]  	& 5537.9 &	6011.0$\pm$1.0 &	-37$\pm$44 &   0.37$\pm$0.08      \\
~[FeVII]	& 5720.9	&  6207.4$\pm$0.5 &   -134$\pm$24 &   0.38$\pm$0.10	 &	\\
~[NII] 	& 5754.6 &	6247.6$\pm$0.2 & 32$\pm$14 	&   0.83$\pm$0.05 	  \\
~HeI & 5875.6 & 	6378.4$\pm$0.2 &	3$\pm$15  &    3.2$\pm$0.03	\\	
~[FeVII]  	& 	6086.9 &	6607.1$\pm$1.8 &   -28$\pm$77 &   0.32$\pm$0.04	 \\
~[OI] 		&	6300.3 	& 6839.4$\pm$0.2 & 1$\pm$15  & 10.1$\pm$0.2	 	  \\
~[SIII]  	& 6312.1 	&	6851.0$\pm$0.3 & -53$\pm$17	& 	1.10$\pm$0.04    \\ 
~[OI]  	&  6363.8 & 6908.5$\pm$0.2 & 7$\pm$15 &   2.89$\pm$0.13   \\
~[NII]  	&	6548.1 & 	7108.7$\pm$0.2 &	14$\pm$15 & 12.4$\pm$1.5	  \\
~H$\alpha$ & 	6562.8  	 & 7124.3$\pm$0.2 &  0$\pm$12 &  100.0 \\ 	
~[NII]  	&	6583.5 & 	7146.8$\pm$0.2 &	-2$\pm$15 & 36.3$\pm$2.6	  \\
~HeI  	& 	6678.2 & 	 7249.7$\pm$0.8 &  4$\pm$36  &  0.92$\pm$0.07   \\		
~[SII] 	&	6716.4  &	 7291.2$\pm$0.2 &	5$\pm$15 &  25.0$\pm$0.7 \\
~[SII]		&      6730.8 		& 7306.7$\pm$0.2 &  6$\pm$15 & 18.8$\pm$0.2 	\\
~HeI		&       7065.7 	& 7670.6$\pm$0.5 & 13$\pm$23  &  0.40$\pm$0.13	 \\
\hline
\end{tabular}
\caption{Emission lines identified in the spectrum of the NE bubble. Columns (2) and (3) give the rest frame and observed $\lambda$. $\Delta(V)$  in (4) is the shift in velocity relative to H$\alpha$. The line ratios in (5) are given relative to H$\alpha$, with $F(H\alpha)$=(7.60$\pm$0.06)$\times$10$^{-15}$ erg s$^{-1}$ cm$^{-2}$.  $^*$Contamination  of [Fe VII]$\lambda$5276.4 by 
 [Fe III] and [FeII] lines is expected to be negligible, given the non detection of other lines by the same species (see Villar Mart\'\i n et al. 2015). }   
\label{tab:bubblelines}
\end{table*}

\section{Discussion}
\label{discuss}

\subsection{The giant nebula}

In the 1980's and early 1990's, long slit spectroscopy and narrow band emission line imaging revealed the existence of emission line gas with typical total sizes $D\la$several 10s of kpc in a significant fraction of radio loud quasars  (Stockton \& Mackenty \citeyear{sto87}) and powerful radio galaxies with strong nuclear emission  (e.g. Dazinger et al. \citeyear{dan84}, Tadhunter et al. \citeyear{tad88}, Baum et al. \citeyear{bau90}). 
Different origins were proposed for the nebulae: tidal debris (remnants of galactic mergers/interactions),   the optical signatures of cooling flows  or gas swept out of the host galaxy by a large-solid-angle blast wave accompanying the production of the radio jets (Stockton \& Mackenty \citeyear{sto87}, Baum et al. \citeyear{bau92}, Fu \& Stockton \citeyear{fu09}). More recent studies have revealed that also radio quiet quasars are associated with extended ionized nebulae (Liu et al. \citeyear{liu13}, Harrison et al. \citeyear{har14}, Humphrey et al. \citeyear{hum15}, Villar Mart\'\i n et al.  \citeyear{vm16}) with typical total extensions  also $D\la$several 10s kpc although it appears that they are frequently smaller than those around radio loud systems. 

 Some ionized nebulae are giant, although     $D\ga$100 kpc   are rare and very extreme at  low $z$. Among the largest   around     active galaxies at $z\la$0.4 are those associated  with\footnote{The sizes have been converted to the same cosmology used in this work}:
 
$\bullet$ the radio galaxies PKS 0349-27 ($z=$0.066, $D\sim$60 kpc, Danziger et al. \citeyear{dan84}), 3C227 ($z$=0.086, $D\sim$111 kpc,  Prieto et al. \citeyear{pri93}),  Coma A ($z$=0.086, $D\sim$70 kpc, van Breugel et al. \citeyear{breu85}, Tadhunter et al.  \citeyear{tad00}), PKS 1932-464  ($z$=0.230, $D\sim$145 kpc, Villar Mart\'\i n et al.  \citeyear{vm05}) 

$\bullet$  the radio loud type 1 quasars 4C37.43  ($z=$0.370, $D\sim$140 kpc, Stockton et al. \citeyear{sto02}), 3CR249.1 ($z=$0.312, $D\sim$67 kpc, Stockton \& MacKenty \citeyear{sto83}), the radio quiet type 1 quasar MR 2251-178 ($z$=0.064, $D\sim$200 kpc, Shopbell et al. \citeyear{sho99})

$\bullet$  the radio quiet type 2 quasars  SDSS J0123+10 ($z=$0.399, $D\sim$180 kpc, Villar Mart\'\i n et al. \citeyear{vm10}), SDSS J1356+1026 ($z=$0.123, $D\sim$80 kpc, Greene et al.  \citeyear{gre12}) and SDSS J1653+23, nicknamed the ``Beetle'' ($z$=0.103, $D\sim$70 kpc, Villar Mart\'\i n et al.  \citeyear{vm17})

  The excitation of the gas is always related at least partially to AGN photoionization.  Stellar photoionization  has also been identified in some of these giant nebulae (e.g. Villar Mart\'\i n et al. \citeyear{vm05},\citeyear{vm10}).  

Giant extensions appear more common in the high $z$  Universe ($z\ga$2), where   $\ga$100 kpc Ly$\alpha$ nebulae are sometimes found
associated with radio loud AGN, both quasars and radio galaxies (McCarthy et al. \citeyear{mcc93}, Heckman et al. \citeyear{hec91},Villar Mart\'\i n \citeyear{vm07}), radio quiet AGN  (Borisova et al. \citeyear{bor16}) and with (apparently) non active galaxies  (Chapman et al. \citeyear{cha01}, Matsuda et al. \citeyear{mat11}). A selection effect  is possibly at work, since more luminous  AGN  can in principle ionize  gas at larger distances. It is   also possible that higher $z$ AGN are  immersed in larger gaseous reservoirs. 

 The nebula we have discovered associated with the ``Teacup'' is  among the largest known at any $z$, specially at $z\la$0.4.  It has two things in common with the low $z$ AGN  with reported giant  nebulae: they are very luminous active galaxies  and they show clear signs of past or ongoing galactic  interactions,  with rich tidal features of different morphologies (shells, tails, bridges, etc).  This is consistent with   \cite{sto02}, who found a correspondence between the incidence of strong extended ionized nebulae among  type 1 quasars and the presence of overt signs of strong galactic interactions, such as close companion galaxies and continuum tails or bridges.
 
  Large-scale (up to almost $\sim$200 kpc) neutral hydrogen HI structures have been detected around several low $z$ radio galaxies (Morganti et al. \citeyear{mor02a},\citeyear{mor02b}, Emonts et al. \citeyear{emo10};  see also Oosterloo et al. \citeyear{oos10})  in emission or in absorption against the radio lobes. They sometimes consist of enormous rotating discs/rings of HI gas, probably resulting from the merger of gas rich galaxies. 
  
All these giant reservoirs of neutral and  ionized gas may be the  manifestation of the same phenomenon:   galactic interactions populate the circumgalactic medium  (CGM) with tidal debris, which sometimes are photoionized and  rendered visible thanks to the illumination by the active nucleus.  
We thus propose that the ``Teacup''  giant nebula is part of the CGM. The optical/radio bubbles, which are comparatively very small, appear to be expanding across this medium. This is suggested by  the observed  impact on the global kinematic pattern in at least some localized positions near or at the bubble edges. 

Some of the gaseous  reservoirs mentioned above have had enough time to settle in  giant rotating disks or rings. 
 The quiescent kinematics of the ``Teacup'' outer nebula is indeed reminiscent of rotation, although integral field spectroscopy would be required to constraint the 2D velocity pattern more accurately. If the  giant nebula is  settled in a  disk with rotational velocity $V_{\rm rot}$, we can estimate  the dynamical mass contained within its radius  $R\sim$50 kpc, via:

$$ M_{\rm dyn}(r<R)\approx \frac{(V_{\rm obs}/sin(i))^2~R}{G}$$

where $i$ is the inclination relative to the plane of the sky ($i$=90$\degr$, edge on), $V_{\rm rot}=\frac{V_{\rm obs}}{sin(i)}$, $V_{\rm obs}\sim$150 km s$^{-1}$ is half the amplitude of the observed rotation curve, as inferred from Fig. \ref{fig:kinem} and $G$ is the gravitational constant.  We have constrained $i\sim$70$\degr$ very roughly using Fig. \ref{fig:hst2}. This implies $ M_{\rm dyn}\sim$3.0$\times$10$^{11}$ M$_{\odot}$. For comparison, the total stellar mass of the ``Teacup'' host galaxy  is  $ M_{\rm stars}\sim$6.3$\times$10$^{10}$ M$_{\odot}$ according to  the Vizier catalogue  (Ochsenbein al. \citeyear{och00})  based on \cite{men14}. Taking into account dark matter, the total mass is probably a few times higher  (Nigoche-Netro et al. \citeyear{nig16}).  Therefore,   although these calculations are based on strong assumptions, in principle the total mass of the ``Teacup'' host galaxy appears sufficient to sustain the rotation of the giant nebula. 

 The emission line spectrum of the giant nebula demonstrates that the AGN can  ionize gas at huge distances ($>$50 kpc) well beyond the optical size of the host galaxy (Sect. \ref{ionization}). Therefore, unless some unknown mechanism protects the molecular reservoir or this is accumulated mostly in a location protected from the hard ionizing continuum,  the quasar can  transform the environment within a huge volume into a very hostile environment for star formation.  It would be interesting to investigate whether this mechanism can explain some of the fundamental issues related to the formation and evolution of massive galaxies that  the community are trying to solve with powerful AGN driven winds (e.g. Silk \& Rees \citeyear{silk98},  King \citeyear{king03}, di Matteo et al. \citeyear{dim05}). Is it possible that the harsh effects of the illumination by active nuclei are enough to quench star formation in the amount we need to solve those fundamental issues?

The CGM is the scene where large scale inflow and outflow from galaxies takes place. The competition between these processes is thought to shape galaxies and drive their evolution. Therefore,  detailed studies of the giant nebulae associated with luminous AGN may probe to be useful to trace back their formation and evolution history. 
The presence of a luminous AGN can render visible the CGM gas around active galaxies which otherwise may not be detected or only  via absorption line studies. 

\subsection{The ionization mechanism and the fading of the AGN}

 We  investigate next whether the AGN can provide sufficient photons to explain the line ratios of the giant nebula. For this we follow \cite{gag14} using our own photoionization models.  
 
 We use the definition of the ionization parameter $U=\frac{Q(H^0)}{4\pi~r^2n~c}$, where $n$ is the gas density, $c$ is the speed of light, $Q(H^0)$ is the  ionizing luminosity emitted by the AGN in photons s$^{-1}$ and $r$ is the distance  in cm from the ionized clouds to the AGN.  $U$ will be constrained with the photoionization models, which together with $r$ and $n$, will provide  $Q(H^0)$.  Because we have very limited information to constrain the models (only [NII]$\lambda$6583/H$\alpha$, [SII]$\lambda \lambda$6716,6731/H$\alpha$ and upper limits on [OI]$\lambda$6300/H$\alpha$, Sect. \ref{ionization}), our goal is to obtain an approximate  $U$ value.   We will then test our results comparing with \cite{gag14} method and results.

We have used sequences of photoionization models built with the code  MAPPINGS1e (Binette et al. \citeyear{bin85},   Ferruit et al. \citeyear{ferr97}) as described in  \cite{sil17} (see also Humphrey et al. \citeyear{hum08}). 
The metallicity  $Z$ ranges from 0.5$Z_{\rm \odot}$ to 2$Z_{\rm \odot}$  in steps of 0.1; the ionization parameter $U$ ranges from  0.0001 to 1.6 in steps of 0.002. A power law of index $\alpha$ is assumed for the ionizing continuum.  Two $\alpha$ values were considered,  -1.0 and -1.5 with  high-energy cut-offs of 10$^3$ eV and 5$\times$10$^4$ eV respectively. For all models, the gas density is $n=$100 cm$^{-3}$. The predicted line ratios were compared with the observed ones and the best model  was selected  using the reduced chi-square ($\chi_{\nu}^2$). 

Given the similar line ratios of Ap. 1 and Ap. 2,  the same optimum model is  selected for both.   If solar abundances $Z=Z_{\rm \odot}$ are considered (as Gagne et al. \citeyear{gag14}), the best model has  $\alpha$=-1.5 and  $U=$0.003 (see Table \ref{tab:tabmodels}, Model I).  For the assumed ionizing continuum,  $\frac{Q_{H^0}}{L_{\rm ion}}\sim$1.6$\times$10$^{10}$, where $L_{\rm ion}$ is the ionizing luminosity in erg s$^{-1}$. Assuming $L_{\rm ion}$=0.35$\times L_{\rm bol}$  (Stern et al. \citeyear{ste14}) we can now infer the bolometric luminosity seen by the gas at the  corresponding distances of Ap. 1 and 2.  We obtain log($L_{\rm bol}$)=46.80 for Ap. 1 and 47.11 for Ap.2. These are $\sim$280 and $\sim$575 times higher than log($L_{\rm bol}$)=44.36 inferred by Gagne et al. (2014)  at the nucleus\footnote{\cite{har15} infer log($L_{\rm bol}$)=45.3 derived from
the midÐfar-infrared spectral energy distribution. We use \cite{gag14} value for a more consistent comparison with their method and results.}. 

 As a test, we have also constrained $L_{\rm bol}$ based on the more sophisticated AGN photoionization models presented by \cite{gag14}, who also assumed solar abundances. We have identified in their Table 4  locations across the bubbles with line ratios as similar as possible to  Ap 1. and Ap. 2 of the giant nebula (e.g. positions (4",0") and (-2",4") using their system of coordinates).  We have then used the $U$ and $n$ values   of the optimum models selected by these authors  for these  locations: log($U_ {\rm high}$)=-2.4, log($n_ {\rm high}$)=2.1; log($U_ {\rm low}$)=-3.2, log($n_ {\rm low}$)=2.9; see their Table 5).  Using the appropriate distances  we now infer as above log($L_{\rm bol}$)$\sim$46.96 for Ap. 1 and 47.28 for Ap.2, which are in  good agreement with our values.     

We find that subsolar abundances  ($Z=0.5Z_{\rm \odot}$)  result on a better agreement between the observed and theoretical ratios\footnote{\cite{keel17} inferred slightly subsolar abundances for the bubbles}. Our best model has $\alpha=$-1.5, $Z$=0.5$Z_{\rm \odot}$  and $U=$0.0006 (Table \ref{tab:tabmodels}, Model II). In this case,   log($L_{\rm bol}$)=46.11 for Ap. 1 and 46.42 for Ap.2, which are $\sim$56 and $\sim$115 times higher than $L_{\rm bol}$ seen by the nuclear gas.

Clearly, the effect of the metallicity on the inferred $L_{\rm bol}$  is important, but both models ($Z_{\rm \odot}$ and 0.5$Z_{\rm \odot}$) imply that the $L_{\rm bol}$ seen by the giant envelope is significantly larger than that seen by gas closer to AGN.

\begin{table*}
\centering
\begin{tabular}{lllllllllll}
\hline
	(1) & (2) & (3) & (4) & (5) \\ 
	Ratio &  Ap .1 & Ap. 2 & Model I & Model II \\ 
		& 18.2 kpc	&	26.2 kpc & $\alpha$=-1.5, $n$=100 cm$^{-3}$ &  $\alpha$=-1.5, $n$=100 cm$^{-3}$  \\
	&	 &	& $Z_{\rm \odot}$, $U$=0.003  &  0.5$Z_{\rm \odot}$, $U$=0.0006  \\  \hline
$\frac{[NII]\lambda 6583}{H\alpha}$ & 0.31$\pm$0.03 &  0.27$\pm$0.03 & 0.36 & 0.29  \\ 
$\frac{[OI]\lambda 6300}{H\alpha}$  & $\la$0.18 & $\la$0.25 &   0.20  &  0.23 \\
$\frac{[SII]\lambda\lambda 6716,6731}{H\alpha}$  & 0.47$\pm$0.04 &  0.40$\pm$0.04  & 0.28 & 0.44 \\ \hline  \hline
 log($L_{\rm bol})$  Ap. 1 &   & & 46.80 & 46.11 & &   \\  
 log($L_{\rm bol})$  Ap. 2 &   & & 47.11 & 46.42 & &   \\  \hline
\end{tabular}
\caption{ Measured line ratios (columns 2 and 3)  for apertures Ap. 1 and  Ap. 2. Columns (4) and (5) show the predicted ratios for the optimum AGN photoionizaton models with solar (4) and half solar (5) abundances. The 0.5$Z_{\rm \odot}$ model reproduces the measured ratios better. The two bottom lines show the bolometric luminosities (log) in erg s$^{-1}$ seen by the gas at the distances of Ap. 1 and 2 implied by   the $U$ values of the  photoionization models.}   
\label{tab:tabmodels}
\end{table*}

\begin{figure}
\includegraphics{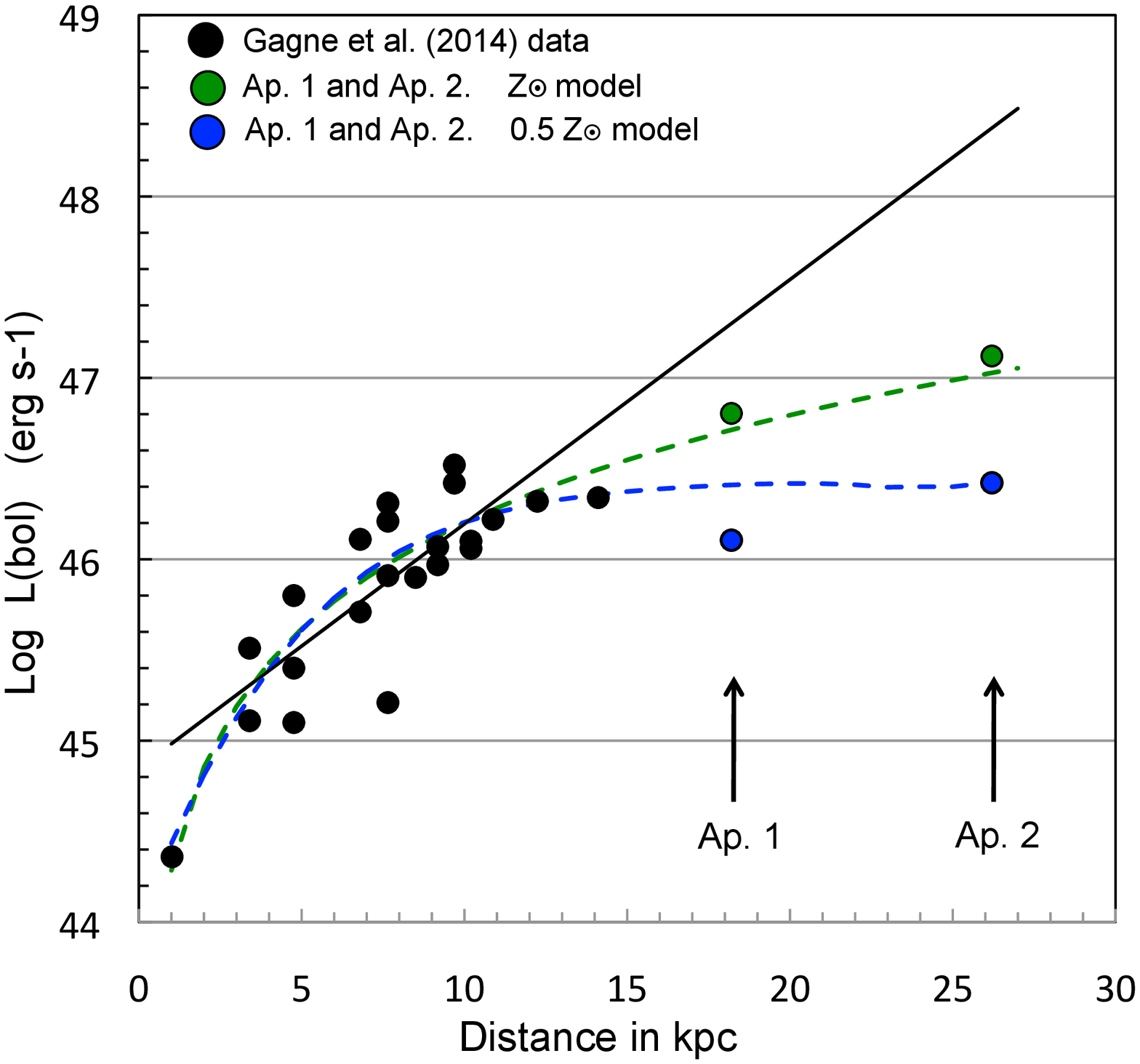}
\vspace{3.1in}
\caption{Fading of the AGN.  Log of bolometric luminosity for the ionized gas at different locations plotted  relative to the projected distance from the AGN. This figure is based on Fig. 11 of Gagne et al. (2014). Ap. 1 and Ap. 2 of the giant nebula are added. The two $L_{\rm bol}$ values  shown for each aperture correspond to the predicted values for our $Z_{\rm \odot}$ (green) and 0.5$Z_{\rm \odot}$ (blue) photoionization models. 
 The linear fit obtained by Gagne et al. (2014)   is shown  in black. The new fits implied by the addition of  the giant nebula Ap. 1 and Ap. 2 are shown in green and blue.  The emission line ratios  of the  nebula imply that  the AGN was more luminous in the past and thus support the quasar fading scenario.}
\label{fig:dimming}
\end{figure}

\cite{rob00} found a related discrepancy for the radio galaxy 3C321. They showed that  the remarkable constancy of the optical line ratios across different spatial regions implies a bolometric luminosity significantly higher than that inferred for the AGN, if the extended  ionized medium consists of an ensemble of ionization bounded clouds. The alternative solution they propose  is that this  medium consists instead of a mixture of optically thin and thick clouds illuminated by a power-law continuum.

 \cite{gag14} propose a different scenario (see also Keel et al. \citeyear{keel17}). These authors  
inferred the  $L_{\rm bol}$ implied by the line ratios measured  at multiple locations across the ``Teacup''  NW and SW bubbles. They found that the $L_{\rm bol}$ seen by the gas decreases with  decreasing distance  to the nucleus, with  a drop of more than two orders of
magnitude from $\sim$15 kpc to the AGN.  They conclude that the active nucleus has decreased in luminosity  in a continuous fashion over $\sim$46,000 yr. 

Our results  are consistent with this scenario   and support that the ``Teacup'' quasar was  more  luminous in the past. We show in  Fig. \ref{fig:dimming} the $L_{\rm bol}$ seen by the gas at increasing distances from the AGN, based in Fig. 11 of \cite{gag14}. Both the solar and specially the subsolar abundance models imply  that the $L_{\rm bol}$  vs. $r$  curve  flattens at increasing distances. This suggests that the dimming rate has not been constant with time. Depending on the true abundance,  the conclusions are different. The $Z_{\rm \odot}$ model  suggests that the fading of the AGN was slower in the past. It  has occurred for $\sim$86,000 yr during which the AGN luminosity  has dropped by $\sim$575 times\footnote{The recombination ($\sim\frac{10^5}{n_e}$ yr) and cooling ($\sim\frac{10^4}{n_e}$ yr) times, where $n_e$ is the electron density in cm$^{-3}$, are short compared to  the light-travel times (Osterbrock \& Ferland \citeyear{ost06}).}. If the giant nebula has subsolar abundances, the dimming started $\sim$46,000 yr ago during which the AGN faded $\sim$100 times  (Gagne et al. \citeyear{gag14}).

 Deep wide field integral field spectroscopy of the giant nebula covering a broad spectral range would be extremely valuable to advance in this study. More optical line ratios (including the critically important [OIII]$\lambda$5007/H$\beta$) measured across the extension of the nebula, up to the measured $R_{\rm max}\sim$50 kpc, would allow to constrain more accurately the  abundance and $U$ values at different locations. This is essential to  characterize more precisely   the temporal variation of the ``Teacup'' bolometric luminosity.

\subsection{Coronal emission from the NE bubble}

In general, the coronal emission in active galaxies is concentrated  within small distances from the AGN,   extending from just a few parsecs  up to 230 pc. This is consistent with the bulk of the coronal lines (CLs) originating between the broad-line region (BLR) and the NLR, and extending into the NLR in the case of  Fe$^{+6}$ and Ne$^{+4}$ lines (e.g. Mazzalay et al. 2010). They are believed to be excited by AGN photoionization (e.g. Mazzalay et al. \citeyear{maz10}). Objects showing coronal emission at such large distances  as the ``Teacup''  ($\sim$10 kpc) are very rare. Another example  is   the radio galaxy PKS 2152-69 at $z=$0.028 (Tadhunter et al. \citeyear{tad88}). A luminous cloud located at $\sim$11 kpc from the nucleus emits CLs, including [FeX]$\lambda$6374.  

The detection of [FeVII] emission along PA60 and PA90 suggests that the mechanism exciting the coronal lines acts across a large volume of the NE bubble. This is against a scenario such that   a highly collimated radio jet excites the lines locally as it interacts  with the gas in its advance through the ambient medium (Tadhunter et al. \citeyear{tad88}).

Based on theoretical arguments, \cite{fer97} showed that CLs form at distances from just outside the BLR (the highest ionization lines) up to a maximum distance  $R_{\rm CL}\sim$400$L_{43.5}^{1/2}$ pc (the lowest ionization lines, such as those emitted by Fe$^{+6}$). $L_{43.5}$ is the ionizing luminosity in units of 10$^{43.5}$ erg s$^{-1}$. This upper limit  could be higher for gas densities $<$100 cm$^{-3}$. 

 Our measured $R_{\rm CL}\sim$9.0 kpc requires $L_{43.5}\sim$506 and this implies log($L_{\rm bol}$)$\sim$46.7.  \cite{gag14} concluded that  the emission line spectrum at this same distance implies  log($L_{\rm bol}$)$\sim$46.0$\pm$0.2, which is remarkably similar, taking into account all possible uncertainties (uncertainty on $R_{\rm CL}$, scatter of the relations $L_{\rm bol}$ vs. Distance for the ``Teacup'', $L_{\rm bol}$ vs. $L_{\rm ion}$ and $R_{\rm CL}$ vs. $L_{43.5}$). For comparison, $R_{\rm CL}\sim$640 pc is inferred for  the current   log($L_{\rm bol}$)=44.36 implied by the nuclear spectrum. It thus appears feasible that  the large size of the coronal region  is consistent with photoionization by the central AGN, provided it was brighter in the past, in consistency  with the fading quasar scenario.

\section{Summary and Conclusions}
\label{Conc}

We present  new results on the radio quiet type 2 quasar (QSO2) nicknamed the ``Teacup'' ($z$=0.085) based on long slit spectroscopic data obtained with the 10.4m Gran Telescopio Canarias (GTC). We have discovered that the QSO2 is associated with a giant reservoir of ionized gas which extends across $\sim$111 kpc along PA60 and at least 71 kpc along PA90. It is among the largest known ionized nebulae associated with  active galaxies at any $z$. The  well known radio/optical   bubbles  ($\sim$10-12 kpc in size) are comparatively tiny. 
  
  We propose that the giant nebula is  part of the circumgalactic medium  of the ``Teacup'', which has been populated with tidal debris (maybe settled in a giant rotating disk) by galactic interactions. This rich gaseous medium has been rendered visible due to the illumination by the powerful active nucleus. The optical/radio bubbles appear to be expanding across this medium, given their impact on the global kinematic pattern in at least some localized positions.

The kinematics of the nebula are much more quiescent (very narrow lines FWHM$\la$150 km s$^{-1}$, flat velocity field at both sides of the AGN) in the outer parts at $\ga$20 kpc from the AGN. The more chaotic motions at smaller radii reveal a different mechanism affecting the kinematics that is in part (possibly not only) related to the  radio/optical bubbles.

The nebula is most likely photoionized by the AGN. Subsolar abundances ($\sim$0.5$Z_{\rm \odot}$) are suggested by the optical line ratios, although a broader spectral coverage would gel to constrain the metallicity more accurately. Its emission line spectrum implies that  the active nucleus was  more luminous in the past, in  consistency with the fading quasar scenario suggested by \cite{gag14}.  Depending on the nebular abundances,  the conclusions are different regarding the temporal variation of the AGN luminosity. The $Z_{\rm \odot}$ model  suggests that the dimming  of the AGN was slower in the past. It  has occurred for $\sim$86,000 yr during which the AGN luminosity  has dropped by $\sim$575 times.  The subsolar abundance model suggests that the dimming started $\sim$46,000 yr ago and the AGN has faded a factor of $\sim$115 since.

We also report the detection of coronal emission (Fe$^{+6}$) from  the NE bubble, at $\sim$9 kpc from the AGN. The detection of coronal lines at such large distances is
very infrequent. For the ``Teacup'' it can be explained by AGN photoionization, provided the QSO2 was much more luminous in the past. This adds further support to   the fading quasar scenario.

 The CGM is the scene where large scale inflow and outflow from galaxies takes place. The competition between these processes is thought to shape galaxies and drive their evolution.  Deep wide field integral field spectroscopy of  powerful AGN with instruments such as MUSE on VLT  opens up a  way to detect and study the elusive material from the CGM around massive active galaxies thanks to the presence of a luminous AGN that can render it visible.

\section*{Acknowledgments}

We thank the referee William Keel for interesting scientific suggestions.

This work is based on observations carried out at the Observatorio Roque de los Muchachos (La Palma, Spain) with  GTC (programme GTC13-16B). We  thank the GTC staff for their support with the observations. 
 MVM and ACL acknowledge  support  from the Spanish Ministerio de Econom\'\i a y Competitividad through the grant AYA2015-64346-C2-2-P. 
 
 The National Radio Astronomy Observatory is a facility of the National Science Foundation operated under cooperative agreement by Associated Universities, Inc.

 This research has made use of  1) the VizieR catalogue access tool, CDS,
 Strasbourg, France. The original description of the VizieR service was
 published in Ochsenbein et al. A\&AS, 143, 23; 2)    the NASA/IPAC circumgalactic Database (NED) which is operated by the Jet Propulsion Laboratory, California Institute of Technology, under contract with the National Aeronautics and Space Administration.

{}

\end{document}